\documentclass[11pt]{article}
\usepackage[latin1]{inputenc} 
\usepackage{amssymb}
\usepackage{amsfonts} 
\usepackage{graphicx} 
\usepackage{amsthm} 
\usepackage{amsmath} 
\usepackage{epsfig}

\usepackage[active]{srcltx}

\def\ni{\noindent}
\def\nn{\nonumber}

\def \bc {\begin{center}}
\def \ec {\end{center}}
\def \bi {\begin{itemize}}
\def \ei {\end{itemize}}

\def \ba {\begin{array}}
\def \ea {\end{array}}

\def \bea {\begin{eqnarray}}
\def \eea {\end{eqnarray}}

\def \be {\begin{equation}}
\def \ee {\end{equation}}

\def \ca {{\cal A}}
\def \cf {{\cal F}}

\def \ct {{\cal T}}
\def \cb {{\cal B}}

\def \cn {{\cal N}}
\def \ch {{\cal H}}

\def \bz {\bar{z}}

\newcommand{\bra}[1]{\langle #1|}
\newcommand{\ket}[1]{|#1\rangle }
\newcommand{\kernel}[2]{\langle #1|#2\rangle}

\newtheorem{thm}{Theorem}[section]
\newtheorem{cor}[thm]{Corollary}
\newtheorem{lem}[thm]{Lemma}
\newtheorem{prop}[thm]{Proposition}
\newtheorem{defn}[thm]{Definition}
\theoremstyle{remark}
\newtheorem{rem}[thm]{Remark}

\newcommand{\li}{\langle}
\newcommand{\ld}{\rangle}

\begin{document}

\begin{center}
{\Large {\bf Almost Complete Coherent State Subsystems and Partial Reconstruction of Wave Functions in the Fock-Bargmann Phase-Number Representation}}
\end{center}
\bigskip

\centerline{{\sc M. Calixto}$^{1}$, {\sc J. Guerrero}$^{2}$
and {\sc J.C. S\'anchez-Monreal}$^{3}$ }

\bigskip

\bc {\it $^1$ Departamento de Matem\'atica Aplicada, Universidad de Granada, Facultad de Ciencias, Campus de Fuentenueva, 18071 Granada, Spain.\,\,\, calixto@ugr.es}
\\
{\it $^2$ Departamento de Matem\'atica Aplicada, Universidad de
Murcia, Facultad de Inform\'atica, Campus de Espinardo, 30100
Murcia, Spain. \,\,\,juguerre@um.es }\\
 {\it $^3$ Departamento de Matem\'atica Aplicada y Estadística, Universidad Politécnica de Cartagena, Paseo Alfonso XIII 56, 30203 Cartagena, Spain.\,\,\, JCarlos.Sanchez@upct.es} \ec


\bigskip
\begin{center}
{\bf Abstract}
\end{center}
\small
\begin{list}{}{\setlength{\leftmargin}{3pc}\setlength{\rightmargin}{3pc}}
\item We provide (partial) reconstruction formulas and discrete Fourier transforms for
wave functions in standard  Fock-Bargmann (holomorphic) phase-number representation from a finite
number $N$ of phase samples $\{\theta_k=2\pi  k/N\}_{k=0}^{N-1}$ for a given mean number $p$ of particles. The resulting
Coherent State (CS) subsystem ${\cal S}=\{|z_k=p^{\frac{1}{2}}e^{i\theta_k}\rangle\}$ is complete (a frame) for truncated Hilbert spaces (finite number of particles) and
reconstruction formulas are exact. For an unbounded number of particles,  ${\cal S}$ is ``almost complete'' (a \textit{pseudo-frame}) and  partial
reconstruction formulas are  provided along with an
study of the accuracy of the approximation, which tends to be exact when $p<N$ and/or $N\to\infty$.

\end{list}
\normalsize 



%
%
%
%
%


\newpage


\section{Introduction}

The subject of completeness of Coherent State (CS) subsystems traces back to von Neumann results concerning the density of canonical CS \cite{Neumann} and
is closely related to standard theorems like Fourier, Heisenberg (uncertainty relations), Nyquist-Shanon (information theory), etc. Criteria of completeness
for a CS subsystem ${\cal S}=\{|z_k\rangle\}$ corresponding to a set of points ${\cal Q}=\{z_k\}$ in the complex plane $\mathbb C$ are known in the literature
(see e.g. the standard text books \cite{CS,Gazeau} and references therein).
In general, ${\cal S}$ is complete iff no function (vector symbol) $\Psi(z)=\langle z|\psi\rangle$, $\psi\not=0$, in the Fock-Bargmann space, vanishing at all
points in ${\cal Q}$, exists. Of special interest is the case where the points $z_k$ form a rectangular lattice $z_k=z_{mn}=m\omega_1+n\omega_2$, where
the periods of the lattice $\omega_1$ and $\omega_2$ are linearly independent, $S\equiv\Im(\bar\omega_2\omega_1)\not=0$, and $m,n$ are arbitrary integers. The simplest
example is the von Neumann square lattice with elementary Planck cell area $S=\pi$, corresponding to the most accurate simultaneous measurement of both coordinate
and momentum. For $S=\pi$, the CS subsystem ${\cal S}=\{|z_{mn}\rangle\}$ is complete and it remains complete if one state is removed, but becomes incomplete when any two states are removed. For
 $S<\pi$, the CS subsystem ${\cal S}$ is overcomplete and remains so if a finite number of states are removed. For  $S>\pi$, the CS subsystem ${\cal S}$ is not complete. This gives a
characterization of overcompleteness in terms of density of states per Planck cell. All these results can  also be stated in the language of (discrete) frames and sampling and
interpolating sets (see for instance \cite{Seip}), and we shall make use
of this terminology in this article.

The square lattice ${\cal Q}=\{z_{mn}\}$ provides a sampling of the (position-momentum) phase space, so that reconstruction formulas for arbitrary functions $\Psi(z)$ can be obtained from
its samples $\Psi(z_{mn})$ for $S\leq \pi$. In this article we shall use an alternative phase-number parametrization of $z=r e^{i\theta}$ in terms of ``mean number of particles'' $p=r^2=|z|^2$ and phase
$\theta=\arctan(\Im(z)/\Re(z))$ and we shall choose a CS subsystem ${\cal S}=\{|z_k\rangle\}$ corresponding to a finite set of $N$ points ${\cal Q}=\{z_k=r e^{2\pi i k/N}\}_{k=0}^{N-1}$ uniformly distributed
on a circle of fixed radius $r$. This would  physically correspond to a finite number $N$ of phases
$\{\theta_k=2\pi i k/N\}_{k=0}^{N-1}$ for a fixed mean number of particles $p=r^2$. Of course, the resulting CS subsystem ${\cal S}=\{|z_k\rangle\}$ won't be complete for finite $N$ but partial (almost exact)
reconstructions of functions  $\Psi(z)$ from $N$ samples $\{\Psi(z_k)\}$ will be possible and we shall study the accuracy of the approximation. We shall also study the conditions under which an exact
reconstruction of $\psi$ is possible, namely when  restricting oneself to a truncated Hilbert subspace (finite number of particles) or for mean number of particles $p$ below the critical value $p_c=N$.
The Fourier coefficients (or probability of occupation numbers) can be
obtained by means of the (filtered) Discrete Fourier transform of the data,
allowing for a straightforward fast extension of the
reconstruction algorithm.

The organization of the paper is as follows.  In Section \ref{CCS} we briefly review some properties of canonical CS  and their restriction to the circle, suggesting the possibility of
reconstructing states from phase samples for a finite mean number of particles. Section \ref{finitenumber} discusses the case of exact reconstruction for truncated (finite-dimensional) Hilbert spaces
(finite number of particles--bosons) in the language of discrete frames, sampling operators and ``sinc-type'' kernels. The more interesting and general case of partial reconstruction
of arbitrary functions in terms of finite (non-complete) CS subsystems (pseudo-frames and undersampling), in the number-phase parametrization, is developed in Section \ref{infinitenumber}.
The advantage of this number-phase parametrization is that it allows the explicit inversion of resolution and reproducing kernel operators by making use of the theory of Circulant Matrices and Rectangular
Fourier Matrices. We also study the accuracy of the approximation and the conditions under which it tends to be exact. Section \ref{comments} is left for some comments on potential
physical applications in Quantum Optics and an outlook. We also
introduce a couple of appendices: Appendix A commenting on traditional problems associated with the proper definition of a phase operator,
and Appendix B with the proof of the main theorem of our work.

\section{Canonical Coherent States and its Restriction to the Circle\label{CCS}}

Bose quanta are described by a pair of annihilation (lowering $a$)
and creation (rising $a^\dag$) ladder operators fulfilling the
commutation relations:
\be [a,a^\dagger]=1.\label{commurel}\ee
The Hilbert (occupation-number or Fock) space ${\cal H}$ realizing
this commutation relation is made of (normalized) eigenstates
$|n\rangle$ of the (Hermitian) number operator $\cn=a^\dag a$,
i.e.:
\be \cn|n\rangle=n|n\rangle, \;\; \langle
n|m\rangle=\delta_{nm}.\ee
These states can be generated from the Fock vacuum  $|0\rangle$
as:
\be|n\rangle=\frac{1}{\sqrt{n!}}(a^\dag)^n|0\rangle\label{Fockbasis}\ee
and they fulfill the completeness condition $\sum_{n=0}^\infty
|n\rangle\langle n|=1$.

Coherent states $|z\rangle, \,z\in \mathbb C$, are defined as
eigenstates of the annihilation operator $a$,
\be a|z\ld=z|z\ld.\ee
They can be explicitly generated by acting with the (unitary)
displacement operator
\be U(z,\bar{z})\equiv e^{z a^\dagger-\bar{z}
a}=e^{-z\bar{z}/2}e^{z a^\dagger}e^{\bar z a},\label{Hopfcomplex}
\ee
on the vacuum $|0\ld$, that is:
\be |z\ld\equiv
U(z,\bz)|0\ld=e^{-z\bar{z}/2}e^{za^\dagger}|0\ld=e^{-z\bar{z}/2}\sum_{n=0}^\infty
\frac{z^n}{\sqrt{n!}}|n\rangle.\label{exp-ansion}\ee

Since $a$ is not hermitian, there is no reason (in principle) for
the set $\{|z\ld, z\in\mathbb C\}$ to span an orthonormal complete
set. In fact, one can easily compute the CS overlap (or
Reproducing Kernel)
\be C(z,z')=\langle
z|z'\rangle= e^{-z\bar{z}/2}e^{-z'\bar{z}'/2}e^{\bar{z}z'},\label{CSoverlap}
\ee
which is not zero for $z\not=z'$. Actually, the set  $\{|z\ld,
z\in\mathbb C\}$ turns out to be overcomplete. In fact, it defines
a \emph{tight frame} in ${\cal H}$, with resolution of unity
\be I=\frac{1}{\pi}\int_{\mathbb C}|z\ld \li
z|d^2z,\label{resolholo}\ee
where we denote $d^2z=d{\rm Re}(z)d{\rm Im}(z)=rdrd\theta$ the
Lebesgue measure in $\mathbb C$ in polar coordinates
$z=re^{i\theta}$. Indeed, using the expansion (\ref{exp-ansion})
we have that
$$ \frac{1}{\pi}\int_{\mathbb C}|z\ld \li
z|d^2z=\frac{1}{\pi}\int_{\mathbb C} e^{-z \bz} 
\sum_{n,m=0}^{\infty}\frac{z^n\bar{z}^m}{\sqrt{n!m!}}\;|n\ld \li
m|=$$$$=\frac{1}{\pi}\sum_{n,m=0}^\infty\int_0^\infty
dr\frac{r^{n+m+1}}{\sqrt{n!m!}}e^{-r^2}\;|n\ld \li
m|\int_0^{2\pi}d\theta
e^{i\theta(n-m)}=$$$$=2\sum_{n=0}^\infty\int_0^\infty
dr\frac{r^{2n+1}}{n!}e^{-r^2}\;|n\ld \li n|=
\sum_{n=0}^\infty|n\ld\li n|=I.$$
The average number of particles $p$ (bosons) in a coherent state
$|z\ld$ is
\be p=\li z|\cn|z\ld=|z|^2\ee
and the probability amplitude of finding $n$ particles in $|z\ld$
is:
\be U_n(z)\equiv\li n|z\ld =\li
n|U(z,\bz)|0\ld=e^{-z\bar{z}/2}\frac{1}{\sqrt{n!}}z^{n}.
\label{upsilon}\ee
In the same way, the probability amplitude of finding any state
 \be |\psi\ld= \sum_{n=0}^{\infty}a_n |n\ld\in {\cal H}\ee
in the coherent state $|z\ld$ is given by:
\be {\Psi}(z)\equiv\li z|\psi\ld=\sum_{n=0}^{\infty}a_n
\overline{U_n(z)}.\ee
The ``Fourier'' coefficients $a_n$ can be calculated through the
following integral formula:
\be a_n=\li n|\psi\ld=\frac{1}{\pi}\int_{\mathbb
C}\Psi(z){U}_n(z)d^2z.\label{CFT}\ee
Due to the over-completeness of the coherent states and to the holomorphic character of the representation, the
integration on the radius $|z|$ is not really necessary. Actually,
we can recover $a_n$ from the values of $\Psi(z)$ on
$z=\sqrt{p}e^{i\theta}$ for a fixed average number of particles
$p=|z|^2$ as:
\be a_n=\sqrt{\frac{n!}{p^{n}e^{-p}}}\frac{1}{2\pi}\int_{-\pi}^\pi
e^{in\theta}\Psi(\sqrt{p}e^{i\theta})d\theta. \label{FourierTransform}\ee
This fact suggests the possibility of reconstructing the state
$\psi$ from a finite number $N$ of phases
$\{\theta_k\}_{k=0}^{N-1}$ and a fixed (but arbitrary) mean number of
particles $p$. We shall pursue this idea in the next Sections and
we shall study the conditions under which total and partial
reconstructions of $\psi$ from $N$ samples $\{\Psi(z_k),
z_k=\sqrt{p}e^{i\theta_k}\}_{k=0}^{N-1}$ are possible and the
accuracy of the approximation for partial reconstructions, which
tends to be exact in the limit $N\to\infty$.

We shall not deal here with the traditional interesting problems associated with the proper definition of phase operators (see e.g. \cite{Dirac,Susskind,Nieto,Pegg1,Pegg2,GazeauPLA,KiblerJPA}). Instead,
we address the reader to Appendix  \ref{phaseopsec} for a short review on the subject. We shall just
point out that the inherent difficulties in defining a proper phase
operator  can be overcome by
restricting oneself to a truncated Hilbert subspace ${\cal H}_M$
of ${\cal H}$ (finite number of particles). One can also circumvent this problem by
considering quantum systems with a large average number of particles (i.e., semi-classical states), for which phase (and number)
uncertainties are negligible.

\section{Finite Number of Particles\label{finitenumber}}

Firstly, we shall work with systems with finite (but arbitrarily large) number of particles $M<\infty$, so that our Hilbert space will be the truncated
state space ${\cal H}_M\subset{\cal H}$ of $M+1$ dimensions. Following the terminology introduced in \cite{Pegg1,Pegg2}, we denote
by physically accessible o preparable states as those that can be obtained from the vacuum state by acting during a finite time
with a source of finite energy and with a finite interaction. These states belong to ${\cal H}_M$ for some integer $M$. There
are also states, denoted by physical states, that are characterized by the fact that $\langle \cn^q\rangle$ is
finite for finite, but arbitrary $q\in\mathbb{N}$. These states include coherent states, squeezed states, thermal equilibrium
states and in general Gaussian states \cite{GaussianStates}.

Before discussing the reconstruction of states $\psi\in {\cal H}_M$ from a finite number of phase samples, we
shall remind some basic definitions on discrete frames, sampling operators and sinc-type kernels. The advantages of using finite tight frames for finite-dimensional Hilbert spaces has been
recently discussed in \cite{GazeauJPArev}.

\subsection{General considerations on discrete frames}

Our purpose is to discretize the integral (\ref{resolholo}), by restricting
ourselves to a discrete subset ${\cal Q}=\{z_k\}\subset \mathbb C$. The question is
whether this restriction will imply a loss of information, that
is, whether the set ${\cal
S}=\{|z_k\rangle, z_k\in{\cal Q}
\}$ constitutes a \textit{discrete frame} itself, with \textit{resolution operator}
\be {\cal A}=\sum_{z_k\in {\cal Q}}|z_k\rangle\langle z_k|. \ee
The operator ${\cal A}$ won't be in general a multiple of the identity.  In fact, a continuous tight frame might contain
discrete non-tight frames, as happens in our case (see later on
Sec. \ref{finitenumber2}).

Let us assume that ${\cal S}$ generates a discrete frame, that is,
there are two positive constants $0<b<B<\infty$ (\emph{frame
bounds}) such that the admissibility condition \be
b\|\psi\|^2\leq\sum_{z_k\in {\cal Q}} |\langle
z_k|\psi\rangle|^2\leq B\|\psi\|^2\label{pbiop2} \ee holds
$\forall \psi\in{\cal H}_M$. To discuss the properties of a frame,
it is convenient to define the frame (or sampling) operator ${\cal
T}:{\cal H}_M\to \ell^2$ given by ${\cal T}(\psi)=\{\langle
z_k|\psi\rangle, \,z_k\in{\cal Q}\}$. Then we can write ${\cal
A}={\cal T}^*{\cal T}$, and the admissibility condition
(\ref{pbiop2}) now adopts the form \be b I\leq {\cal T}^*{\cal
T}\leq B I,\ee where $I$ denotes the identity operator in ${\cal
H}_M$. This implies that ${\cal A}$ is invertible. If we define the
\emph{dual frame} $\{|\tilde{z}\rangle\equiv {\cal A}^{-1}
|z\rangle\}$, one can easily prove that the expansion
(\emph{reconstruction formula})
\be |\psi\rangle=\sum_{z_k\in {\cal Q}}
\Psi_k|\tilde{z}_k\rangle,\label{reconstructionformula-over}\ee
where $\Psi_k\equiv \langle z_k|\psi\rangle$, holds
in ${\cal H}_M$, that is, the expression \be{\cal T}_l^+{\cal T}=
\sum_{z_k\in {\cal Q}}|\tilde{z}_k \rangle\langle z_k|= {{\cal
T}}^*({\cal T}_l^+)^*= \sum_{{z}_k\in {\cal Q}}|z_k \rangle\langle
\tilde{z}_k |= I\label{resolucionidentidad}\ee
provides a resolution of the identity, where ${\cal
T}_l^+\equiv({\cal T}^*{\cal T})^{-1}{\cal T}^*$ is the (left)
pseudoinverse (see, for instance, \cite{pseudoinverse}) of ${\cal
T}$ (see e.g. \cite{Gazeau,Holschneider} for a proof, where the authors
introduce the dual frame operator $\tilde{\ct}=(\ct_l^+)^*$
instead).

It is interesting to note that the operator $P={\cal T}{\cal
T}_l^+$ acting on $\ell^2$ is an orthogonal projector onto the
range of $\ct$. Eventually, we are interested in a finite number $N$ of phase samples, so that the space $\ell^2$ can be substituted by $\mathbb{C}^{N}$, with $N$ the cardinal of ${\cal Q}$.

{}From (\ref{reconstructionformula-over}) the function $\Psi(z)$
can be obtained
\be \Psi(z)\equiv \langle z|\psi\rangle = \sum_{z_k\in {\cal Q}}
\Xi_k(z)\Psi_k \ee
from its samples $\Psi_k=\langle z_k|\psi\rangle$, through some
``sinc-type'' kernel
\be \Xi_k(z)=\langle z|\tilde{z}_k\rangle\label{sinctype} \ee
fulfilling $\Xi_k(z_l)=P_{lk}$. A projector is obtained, instead
of the identity, to account for the fact that an arbitrary set of
overcomplete data $\Psi_k\in \mathbb{C}^{N}$, can be incompatible with
$|\psi\rangle\in {\cal H}_M$, and therefore they are previously
projected (note that an overdetermined system of equations is
being solved).

This case will be named \emph{oversampling}, since there are more
data than unknowns, and will be discussed in the following  Subsection. In other
contexts, when eq.  (\ref{pbiop2}) holds, the set ${\cal Q}$ is
said to be \emph{sampling} for the space ${\cal H}_M$ \cite{Fuhr}.


\subsection{Exact reconstruction of truncated states}\label{finitenumber2}

In our case, there is a convenient way to select the sampling points $z_k$ in such a way that
 the resolution operator ${\cal A}$ is invertible
 and  explicit formulas for their inverses
are available. These are given by the set of $N$ points (phases) uniformly distributed on a circumference of
radius $\sqrt{p}$:
\be{\cal Q}=\{z_k=\sqrt{p} e^{2\pi i k/N}, k=0,1,\dots N-1\},\label{discreteQ}\ee
Denote by ${\cal S}=\{|z_k\rangle\,,k=0,1,\ldots,N-1\}$ the CS subsystem associated with
the points in ${\cal Q}$ and by
\be {\cal H}^{\cal S}\equiv  {\rm Span}(|z_0\rangle,|z_1\rangle,\dots,|z_{N-1}\rangle)\label{HsS} \ee
the subspace of ${\cal H}$ spanned by ${\cal S}$. For finite $N$ we have ${\cal H}^{\cal S}\not={\cal
H}$, so that we cannot reconstruct exactly every function $\psi\in {\cal H}$ from $N$ of its samples
$\Psi_k=\langle z_k|\psi\rangle$, but we shall proof that for $\psi\in {\cal H}_M$ we can always provide an exact reconstruction formula.

{\defn \label{bandlimited} We define the subspace ${\cal H}_M$ of truncated functions with a finite number of particles
$M<\infty$ as:
\be  {\cal H}_M\equiv {\rm Span}(|0\rangle,|1\rangle,\dots,|M\rangle).\label{band-lim-space}\ee
} The subspace ${\cal H}_M$ is clearly a finite ($M+1$)-dimensional vector subspace of ${\cal H}$.

{\thm \label{maintheorem-over} Given a truncated function $\psi\in {\cal H}_M$ on  $\mathbb
C$, with a finite expansion
\be |\psi\rangle=\sum_{m=0}^M a_m|m\rangle,\label{FourExpan}\ee
there exists a reconstruction formula (\ref{reconstructionformula-over}) of $\psi$
\be {\Psi}(z)=\sum_{k=0}^{N-1} \Xi_k(z)\Psi_k,\label{reconstruccionover} \ee
from $N>M$ of its samples $\Psi_k$ taken at the sampling points (phases) in (\ref{discreteQ}), through a
``sinc-type'' kernel  given by
\be \Xi_k(z)=\frac{1}{N}e^{(z_k\bar{z}_k-z\bar{z})/2}\sum_{m=0}^M(\overline{z
z_k^{-1}})^{m}.\label{xiover}\ee }
Firstly, we shall introduce some notation and prove some previous lemmas.

{\lem \label{lemmaover1}  For $N>M$, the frame operator $\ct:{\cal H}_M\to \mathbb C^N$ given by
$\ct(\psi)=\{\langle z_k|\psi\rangle, z_k\in {\cal Q}\}$ [remember the construction after Eq.
(\ref{pbiop2})] is such that the resolution operator ${\cal A}=\ct^*\ct$ is diagonal, ${\cal A}={\rm
diag}(\lambda_0,\ldots,\lambda_{M})$, in the basis (\ref{band-lim-space}) of ${\cal H}_M$,
with\footnote{The quantities $\lambda_m$ are well defined for $m\in \mathbb{N}\cup \{0\}$ and they will
be used in the case of an infinite number of particles $M=\infty$ too. Note also that $f(n;p)=\lambda_n(p)/N$
is the density function of the Poisson distribution.}
\be \lambda_m(p)\equiv N\frac{e^{-p}}{m!}p^{m},\,m=0,\dots,M.\label{lambdaover}\ee
Hence, ${\cal A}$ is invertible in ${\cal H}_M$. Therefore, denoting $|\tilde{z}_k\rangle\equiv {\cal
A}^{-1}|z_k\rangle$, the dual frame, the expression
\be I_{M}=\sum_{k=0}^{N-1}|{z}_k\rangle \langle \tilde{z}_k| = \sum_{k=0}^{N-1}|\tilde{z}_k\rangle
\langle {z}_k | \label{resolident}\ee
provides a resolution of the identity in ${\cal H}_M$. \label{lemaover}}

\ni \textbf{Proof.} Using (\ref{upsilon}), the matix elements of $\cal T$ can be written as:
\be \ct_{kn}\equiv\kernel{z_k}{n}=e^{-p/2}\frac{p^{n/2}}{\sqrt{n!}}e^{-i2\pi
kn/N}\equiv\lambda^{1/2}_n\cf_{kn}\label{calT}\ee
where $\cf$ denotes the Rectangular Fourier Matrix (RFM)  \cite{samplingsphere}   given by $\cf_{kn}=\frac{1}{\sqrt{N}}e^{-i2\pi kn/N},
k=0,\ldots,N-1,\;n=0,\ldots,M$. Then, the matrix elements of the resolution operator turn out to be:
\be\ca_{nm}\equiv\bra{n}\ca\ket{m}=\sum_{k=0}^{N-1}\ct_{kn}^\ast\ct_{km}=(\lambda_n\lambda_m)^{1/2}\sum_{k=0}^{N-1}\cf_{kn}^\ast\cf_{km}=
\lambda_n\delta_{nm},\ee
where we have used the well known orthogonality relation for RFM:
\bea N\sum_{k=0}^{N-1}{\cal F}^*_{nk}{\cal
F}_{km}&=&\sum_{k=0}^{N-1}\left( e^{2\pi
i(n-m)/N}\right)^k=\left\{\ba{l} N,\;{\rm if}\; (n-m) \,{\rm
mod}\, N = 0\, \\ 0,\;{\rm if}\; (n-m)\,{\rm mod} \,N \not=0
\,\,\ea\right\}\nn\\ &=&N \delta_{(n-m) \,{\rm mod}
N,0},\label{exponencial}\eea
and this equals $N\delta_{n,m}$ if $N>M$. Since ${\cal A}$ is
diagonal with
non-zero diagonal elements $\lambda_n$, then it is invertible and
a dual frame and a (left) pseudoinverse for $\ct$ can be
constructed, ${\cal T}_l^+\equiv \ca^{-1}{\cal T}^*$, providing,
according to eq. (\ref{resolucionidentidad}), a resolution of the
identity. $\blacksquare$
\\
\ni \textbf{Proof of Theorem \ref{maintheorem-over}.} From the resolution of the identity
(\ref{resolident}), any $\psi\in {\cal H}_M$ can be written as $|\psi\rangle = \sum_{k=0}^{N-1}\Psi_k
|\tilde{z}_k\rangle$, and therefore $\Psi(z)=\langle z|\psi\rangle= \sum_{k=0}^{N-1}\Psi_k\langle
z|\tilde{z}_k\rangle$. Using that $|\tilde{z}_k\rangle = \ca^{-1}|z_k\rangle$, we derive that
\bea \langle
z|\tilde{z}_k\rangle &=& \sum_{m,n=0}^M\kernel{z}{m}\bra{m}\ca^{-1}\ket{n}\kernel{n}{z_k}=
\sum_{m,n=0}^M\kernel{z}{m}\ca^{-1}_{mn}\ct_{kn}^\ast\nn =
\sum_{m,n=0}^M\kernel{z}{m}\lambda^{-1}_{n}\delta_{nm}\ct_{kn}^\ast\\
&=& \sum_{m=0}^M\kernel{z}{m}\lambda^{-1}_{m}\ct_{km}^\ast=\sum_{m=0}^M\kernel{z}{m}\lambda^{-1}_{m}
\lambda_m^{1/2}\cf_{km}^\ast\nn=\frac{1}{\sqrt{N}}\sum_{m=0}^M\kernel{z}{m}\lambda^{-1/2}_{m}e^{i2\pi
km/N}\\
&=&\frac{1}{N}\sum_{m=0}^Me^{-z\bar{z}/2}\bar{z}^me^{p/2}p^{-m/2}e^{i2\pi km/N}=
\frac{1}{N}e^{(z_k\bar{z}_k-z\bar{z})/2}\sum_{m=0}^M(\overline{z z_k^{-1}})^{m}= \Xi_k(z),\eea
where we have used the expressions of  $\langle z|m\rangle$ given by eq. (\ref{upsilon}) and eq.
(\ref{discreteQ}), and that $p=z_k\bar{z_k}$. $\blacksquare$

{\rem \label{lagrange} It is interesting to note that eq. (\ref{reconstruccionover}) can be interpreted
as a sinc-type reconstruction formula, where the role of the sinc function  are played by the functions
$\Xi_k(z)$, satisfying the ``orthogonality relations"
\be
\Xi_k(z_l)=\frac{1}{N}\sum_{m=0}^Me^{2\pi i(k-l)m/N}=P_{lk},\ee
 with $P=\ct
\ct_l^+$ an orthogonal projector onto a $M$-dimensional subspace of $\mathbb{C}^N$, the range of $\ct$. In the
case of critical sampling, $N=M+1$, the result $\Xi_k(z_l)=\delta_{lk}$
is recovered (corresponding to
an interpolation formula), but for the strict oversampling case, $N>M+1$, a projector is obtained to
account for the fact that an arbitrary set of overcomplete data $\Psi_k,\,k=0,\ldots,N-1$, can be
incompatible with $|\psi\rangle\in {\cal H}_M$.$\square$
}

A reconstruction in terms of the Fourier coefficients can be directly obtained by means of the (left)
pseudoinverse of the frame operator $\ct$:

{\cor \label{corolarioover} {\rm (Discrete Fourier Transform)} The Fourier coefficients $a_m$ of the
expansion $|\psi\rangle = \sum_{m=0}^M a_m|m\rangle$  of any $\psi\in{\cal H}_M$  can be determined in
terms of the data $\Psi_k=\langle z_k|\psi\rangle$ as
\be a_{m}=\frac{1}{\sqrt{N\lambda_m}} \sum_{k=0}^{N-1}e^{2\pi ikm/N}\Psi_k \,,\,m=0,\ldots,M\,.
\label{coeffourierover} \ee }

\ni \textbf{Proof.} Taking the scalar product with $\langle z_k|$ in the expression (\ref{FourExpan}) of
$|\psi\rangle$, we arrive at the over-determined system of equations
\be \sum_{m=0}^{M} \ct_{km}a_{m}=\Psi_k,\;\;\ct_{km}=\langle z_k|m\rangle, \label{sistema} \ee
which can be solved by left multiplying it by the (left) pseudoinverse of $\ct$,
$\ct_l^+=(\ct^*\ct)^{-1}\ct^*=\ca^{-1}\ct^*$. Using the expressions of $\ca^{-1}={\rm
diag}(\lambda_0^{-1},\lambda_1^{-1},\ldots,\lambda_{M}^{-1})$, given in Lemma \ref{lemmaover1}, and the
matrix elements $\ct_{kn}$, given by the formula (\ref{upsilon}), we arrive at the desired
result.$\blacksquare$

{\rem Note that the Fourier coefficients $a_m$ are obtained as a (rectangular) discrete Fourier
transform of the data $\Psi(z_k)$ up to a normalizing factor  $1/\sqrt{\lambda_m}$ (that is, a
modulation by $\ca^{-1/2}$), which can be seen as a filter. The expression (\ref{coeffourierover})
provides a discretization of (\ref{FourierTransform}), admitting an extension to a fast Fourier Transform.}

\section{Infinite Number of Particles\label{infinitenumber}}

In this section we study the more interesting/realistic case of states with an indeterminate (infinite) number of particles.
Among these we can find the physical states commented in the previous section, for which many of the results presented here
simplify due to the finiteness of all number operator moments.

Before studying the partial reconstruction of
states from a finite number of phase samples, let us introduce some general definitions.

\subsection{Discrete pseudo-frames and pseudo-inverses}\label{CSandFrames}
We shall be mainly interested in cases where there are not enough
phase samples to completely reconstruct a given function $\psi$. We shall call this case
\emph{undersampling}. However, a partial reconstruction is still
possible. In these cases, ${\cal S}$ does not generate a discrete
frame but a ``pseudo-frame'', and the resolution operator ${\cal A}$ would not be
invertible. But we can construct another operator from ${\cal T}$,
${\cal B}\equiv{\cal T}{\cal T}^*$, acting on $\ell^2$. The matrix elements of ${\cal B}$ are
\be {\cal B}_{kl}=\langle z_k|z_l\rangle\,,\label{overlapping} \ee
therefore ${\cal B}$ is the discrete reproducing kernel operator,
see Eq. (\ref{CSoverlap}). If the set ${\cal S}$ is linearly
independent, the operator ${\cal B}$ will be invertible and a
(right) pseudoinverse can be constructed for ${\cal T}$, ${\cal
T}_r^+\equiv {\cal T}^*({\cal T}{\cal T}^*)^{-1}$, in such a way
that $\ct \ct_r^+ = I_{\ell^2}$. As in the previous case there is
another operator, $P_{\cal S}= \ct_r^+ \ct$ acting on ${\cal H}$
which is an orthogonal projector onto the subspace ${\cal H}^{\cal
S}$ spanned by ${\cal S}$. A pseudo-dual frame can be defined as
\be |\tilde{z}_k\rangle = \sum_{z_l\in {\cal Q}} ({\cal
B}^{-1})_{lk}|z_l\rangle \label{discrete-repker}\ee
providing a resolution of the projector $P_{\cal S}$,
\be
 {\cal T}_r^+ \ct = \sum_{z_k\in {\cal Q}}|\tilde{z}_k \rangle\langle z_k|=
\ct^* ({\cal T}_r^+)^*  = \sum_{{z}_k\in {\cal Q}}|z_k
\rangle\langle \tilde{z}_k |= P_{\cal S}
\label{resolucionproyector} \ee
Using this, a partial reconstruction (an ``alias'') $\hat{\psi}$
of $\psi$ is obtained,
\be \hat{\Psi}(z)=\langle z| \hat{\psi}\rangle=\sum_{z_k\in {\cal
Q}} L_k(z)\Psi_k,\label{partialrec} \ee
from its samples $\Psi_k=\langle z_k|\psi\rangle$, through some
 ``Lagrange-like'' interpolating functions
\be L_k(z)=\langle z|\tilde{z}_k\rangle\label{interpolating} \ee
fulfilling $L_k(z_l)=\delta_{kl}$. The alias $\hat{\psi}$ is the
orthogonal projection of $\psi$ onto the subspace ${\cal H}^{\cal
S}$, that is, $|\hat{\psi}\rangle=P_{\cal S}|\psi\rangle$. The
relative (normalized) distance from the exact $\psi$ to the
reconstructed function $\hat{\psi}$ is given by the relative error
function:
\be {\cal E}_\psi({\cal H}^{\cal S})=\frac{\|
\psi-\hat{\psi}\|}{\|\psi\|}=\sqrt{\frac{\langle\psi|I-P_{\cal
S}|\psi\rangle}{\langle \psi|\psi\rangle}}\label{errorpsi}\ee

As mentioned above, we shall denote this case by
\emph{undersampling}, since there are not enough data to fully
reconstruct $\psi$. In other contexts, a set ${\cal Q}$ is said to
be \emph{interpolating} if, for an arbitrary set of data $\Psi_k$
there exists a $|\psi\rangle\in {\cal H}$ such that $\langle
z_k|\psi\rangle = \Psi_k$ \cite{Fuhr}. This condition is satisfied
in this case since $L_k(z_l)=\delta_{kl}$.

The two operators ${\cal A}$ and ${\cal B}$ are intertwined by the
frame operator ${\cal T}$, ${\cal T}{\cal A}={\cal B}{\cal T}$. If
${\cal T}$ were invertible, then both ${\cal A}$ and ${\cal B}$
would be invertible and ${\cal T}_r^+={\cal T}_l^+ ={\cal
T}^{-1}$. This case would correspond to \emph{critical sampling},
where both operators ${\cal A}$ and ${\cal B}$ can be used to
fully reconstruct the function $\psi$. However, in many cases it
is not possible to find a set of points ${\cal Q}$ such that both
${\cal A}$ and ${\cal B}$ are invertible, that is, there is no
critical sampling, or there are not sets ${\cal Q}$ which are
sampling and interpolating at the same time. The most common
example is the Bargmann-Fock space of analytical functions on
$\mathbb{C}$, where one can find rectangular lattices which are
sampling (and therefore ${\cal A}$ is invertible), or which are
interpolating (and thus ${\cal B}$ is invertible), but not both
simultaneously \cite{CS,Fuhr}. Examples of critical sampling are
given by the space of band limited functions on $\mathbb{R}$ and
the set $\mathbb{Z}$, which is both sampling and interpolating,
and the space of functions on the Riemann sphere (or rather its
stereographic projection onto the complex plane) with fixed
angular momentum $s$ and the set of $N^{\rm th}$-roots of unity,
with $N=2s+1$ \cite{samplingsphere}.

It should be noted that in the case in which there is a finite
number $N$ of sampling points $z_k$, the space $\ell^2$ should be
substituted by $\mathbb{C}^N$, and the operator $\cb$ can be
identified with its matrix once a basis has been chosen.


\subsection{Partial reconstruction of states and aliasing\label{partialsec}}

In the previous Section we have seen that, using $N$ phase samples, we can fully
reconstruct functions $\psi\in {\cal H}_M$ for a number of particles up to $M=N-1$. When the
reconstruction of a function  $\psi\in {\cal H}$  from a finite
number $N$ of samples is required, we cannot use the results of the previous section since the
resolution operator $\ca$ is no longer invertible\footnote{While the operator $\ct:\ch\rightarrow
\mathbb{C}^N$ has the same expression as in the previous section, the operator $\ca$ is an infinite
dimensional matrix given by: $\ca_{mn}=\lambda_{j+pN}^{1/2}\lambda_{j'+qN}^{1/2}\delta_{jj'}$, with
$m=j+pN$ and $n=j'+qN$, that is, it is a matrix made of $N\times N$ diagonal blocks.}. However, the
overlapping kernel operator $\cb$ turns out to be invertible, and a partial reconstruction can be done
following the guidelines of the previous  Sec. \ref{CSandFrames} (undersampling).

The space $\ch$ is infinite-dimensional, and therefore in the partial reconstruction of an arbitrary function
$|\psi\rangle=\sum_{n=0}^\infty a_n|n\rangle$ a considerable error will be committed unless further
assumptions on the probability amplitudes, $a_n$, are made. Since $|\psi\rangle$ is normalizable, the
Fourier coefficients $a_n$ should decrease to zero, thus even if $|\psi\rangle\not\in{\cal H}_M $, if $a_n$
goes to zero fast enough, it will ``approximately'' belong   to ${\cal H}_M $  if the norm of
$|\psi_M^\perp\rangle\equiv \sum_{n=M+1}^\infty a_n|n\rangle$ is small compared to  the norm of
$|\psi\rangle$, for an appropriately chosen $M$. Let us formally state these ideas.

{\defn Let us define by
\be P_M=\sum_{m=0}^M |m\rangle \langle m|\ee
the projector onto the truncated space  ${\cal H}_M$ for some $M<\infty$. We shall denote
by
\be \epsilon_{M+1}^2(\psi)\equiv {\cal E}_\psi^2({\cal H}_M)=
\frac{\langle\psi|I-P_M|\psi\rangle}{\langle\psi|\psi\rangle}=
\frac{\sum_{n=M+1}^\infty|a_n|^2}{\sum_{n=0}^\infty|a_n|^2}\,.\label{quasi-bandlimited}\ee
the normalized squared distance [similar to (\ref{errorpsi})] from
\be |\psi\rangle=\sum_{n=0}^\infty a_n|n\rangle \in {\cal H},\label{FourExpan2}\ee
to its orthogonal projection
\be |\psi_M\rangle=P_M|\psi\rangle=\sum_{n=0}^M a_n|n\rangle \label{FourExpan3}\ee
onto the subspace ${\cal H}_M$. In other words, $\epsilon_{M+1}(\psi)$ is the sine of the angle
between $\psi$ and $\psi_M$.}

We hope that the (normalized) error committed when reconstructing
$\psi$ from $N$ of its samples $\Psi_k$ will be of the order of
$\epsilon_{N}(\psi)$ (here $M=N-1$), which will be small as long as the Fourier
coefficients $a_n$ decay fast enough. More precisely, if
$|a_n|\leq C/n^\alpha$ for some constant $C$, $\alpha> 1/2$ and
$n\geq N$, then
\be\|\psi\|^2\epsilon_N^2(\psi)=\sum_{n=N}^\infty|a_n|^2\leq
C^2\sum_{n=N}^\infty\frac{1}{n^{2\alpha}}\leq
C^2\int_{N-1}^\infty\frac{1}{x^{2\alpha}}dx\leq
\frac{C^2}{2\alpha-1}\frac{1}{(N-1)^{2\alpha-1}},\label{asympepsilonN}\ee
which says that $\epsilon_N^2(\psi)=O(\frac{1}{N^{2\alpha-1}})$.\footnote{
This condition could be more formally stated by saying that $\psi$
belongs to a certain Sobolev space $\mathbb H^k$ with
$k<\alpha-1/2$. }

For the special case in which $|\psi\rangle$ is a physical state, for instance a coherent state $|\zeta\rangle$, i.e. when $a_n=e^{-|\zeta|^2/2}\frac{\zeta^n}{\sqrt{n!}}$, we have that
$\epsilon_N^2(\zeta)=1-\Gamma(N,|\zeta|^2)/\Gamma(N)$, which approximately verifies for large $N$:
\be
\epsilon_N^2(\zeta)\simeq \left\{\ba{lcr} 0& {\rm if} & |\zeta|^2<N\\ 1& {\rm if} & |\zeta|^2>N\ea\right..
\ee
This means that, taking $M=N-1$ and denoting by $p=|\zeta|^2$ the mean number of particles, if $N>p$ then
$|\zeta_M \rangle\simeq |\zeta\rangle$, whereas if $N<p$ then $|\zeta_M \rangle\simeq 0$. This behavior is similar
for physical states, and guarantees that the partial reconstruction will be rather accurate.
See later on Eq. (\ref{Heaviside}) for a related discussion on this ``droplet'' behavior.

In the next theorem we provide a partial
reconstruction formula for general functions $\psi\in{\cal H}$ and a bound
for the error committed.

 {\thm \label{maintheoremunder} Given  $\psi\in {\cal H}$, there exists a partial reconstruction of $\psi$, in terms of
the alias
\be \hat{\Psi}(z)=\sum_{k=0}^{N-1}
L_k(z)\Psi_k,\label{reconstruccionunder} \ee
from $N$ of its samples $\Psi_k$, taken at the sampling points in
(\ref{discreteQ}), up to an error (\ref{errorpsi})
\be {\cal E}_\psi^2({p},N)\equiv {\cal E}_\psi^2({\cal H}^{\cal S}) \leq
\frac{\nu_0(p,N)}{1+\nu_0(p,N)}+2\epsilon_{N}(\psi)\sqrt{1-\epsilon_{N}^2(\psi)}+
\epsilon_{N}^2(\psi)\frac{2+\nu_0({p},N)}{1+\nu_0({p},N)}, \label{errorpsi2}\ee
with $\nu_0({p},N)\equiv
\sum_{u=1}^{\infty}\frac{1}{(uN)!}p^{uN}$. The Lagrange-like interpolating
functions (\ref{interpolating}) now adopt the following form:
\be L_k(z)=\frac{1}{N}e^{(z_k\bar{z}_k-z\bar{z})/2}\sum_{j=0}^{N-1}\hat{\lambda}_j^{-1}\sum_{q=0}^\infty
\lambda_{j+qN}\,(\overline{z z_k^{-1}})^{j+qN},\label{hatinterpolating}\ee
where
\be \hat{\lambda}_j=\sum_{q=0}^\infty
\lambda_{j+qN},\,j=0,\dots,N-1,\label{lambdagorro}\ee
are the eigenvalues of the discrete reproducing kernel operator
${\cal B}={\cal T}{\cal T}^*$ [defined in (\ref{overlapping}) with
matrix elements ${\cal B}_{kl}=\langle z_k|z_l\rangle$] and
$\lambda_n$ is given by (\ref{lambdaover}), but now for
$n=0,1,2,\ldots $\,. }

The proof of this theorem is quite involved and we move it to the Appendix \ref{AppendixB} in order to make 
the presentation more dynamic. There we also prove an asymptotic expression  (\ref{varepsilon0asymp}) for $\nu_0({p},N)$, so that 
the quadratic error (\ref{errorpsi2}) approaches zero when
$N\to\infty$, with the asymptotic behavior
\be {\cal E}_\psi^2(p,N)\leq
2\epsilon_{N}(\psi)\sqrt{1-\epsilon_{N}^2(\psi)}+
2\epsilon_{N}^2(\psi)+O(N^{-N}), \label{errorpsiasymp}\ee
as long as $p<p_0(N)$ given in (\ref{roptimo}). Thus, the reconstruction of $\psi$ by $\hat\psi$ is
exact in this limit.



The Fourier coefficients of the alias $\hat{\psi}$ can also be computed, providing a discrete version of the Fourier Transform
(\ref{FourierTransform}), which also admits a fast version.

{\cor {\rm (Discrete Fourier Transform)} The Fourier coefficients
$a_n$ of the expansion (\ref{FourExpan2}) can be approximated by
the discrete Fourier transform:
\be \hat{a}_n=\frac{\lambda_n^{1/2}}{\hat{\lambda}_{n \,{\rm mod}\,
N}} \frac{1}{\sqrt{N}}\sum_{k=0}^{N-1}e^{2\pi i
nk/N}\Psi_k,\label{dftunder}\ee
}

\ni \textbf{Proof.} The Fourier coefficients of the alias
(\ref{reconstruccionunder}) are given by:
 \be
 \hat{a}_n = \langle s,n|\hat{\psi}\rangle= \langle s,n|P_{\cal S}|\psi\rangle= \sum_{k=0}^{N-1}\langle s,n|\tilde
z_k\rangle\Psi_k=\sum_{k,l=0}^{N-1}\langle s,n|
z_l\rangle(\cb^{-1})_{kl}\Psi_k \,.\ee
where $P_{\cal S}$ is defined in Appendix B. Using that ${\cal T}_{ln}=\langle
z_l|s,n\rangle=\lambda_n^{1/2}{\cal F}_{ln}$, given in
(\ref{calT}), and the expression for the inverse of ${\cal B}$
given in (\ref{inverseB}), we obtain the final result once the
orthogonality relation (\ref{exponencial}) for Fourier Matrices is
used. $\blacksquare$

{\rem The expression for the Fourier coefficients $\hat{a}_n$
entails a kind of ``periodization" of the original $a_n$. Indeed,
putting
\[\Psi_k=\langle z_k|\psi\rangle=\sum_{m=0}^\infty \langle
z_k|m\rangle\langle m|\psi\rangle=\sum_{m=0}^\infty {\cal
T}_{km}a_m=\sum_{m=0}^\infty \lambda_m^{1/2}{\cal F}_{km}a_m\]
and using the orthogonality relations (\ref{exponencial}) we can
write (\ref{dftunder}) as:
\bea\hat{a}_n&=&\frac{\lambda_n^{1/2}}{\hat{\lambda}_{n {\rm mod}
N}} \sum_{k=0}^{N-1}{\cal F}_{nk}^*\sum_{m=0}^\infty \lambda_m^{1/2}{\cal F}_{km}a_m\nn\\
&=&\frac{\lambda_n^{1/2}}{\hat{\lambda}_{j}} \sum_{q=0}^\infty
\lambda^{1/2}_{j+qN}a_{j+qN},\;j=n\, {\rm mod} \,N,\eea
which implies
\be
\lambda_n^{-1/2}\hat{a}_n=\lambda_{n+pN}^{-1/2}\hat{a}_{n+pN}\Rightarrow
\hat{a}_{n+pN}=\sqrt{\frac{\lambda_{n+pN}}{\lambda_n}}\,
\hat{a}_n\,,\,\forall p\in\mathbb{N}\,. \ee
This is the ``complex plane'' counterpart of the typical
\emph{aliasing effect} for band-unlimited signals on the real
line.$\square$}

We could think that, for the case $\epsilon_{N}(\psi)=0$, we
should recover the results of Section \ref{finitenumber2}, but we
shall see that this is not the case. Before, a process of
truncation and filtering of $|\hat{\psi}\rangle$ in
(\ref{reconstruccionunder}) is necessary to recover the
reconstruction formula (\ref{reconstruccionover}) for functions $\psi\in {\cal H}_M$ (\ref{FourExpan}). Indeed, if $M=N-1$, the
truncation operation
\be |\hat{\psi}_M\rangle\equiv
P_M|\hat{\psi}\rangle=\sum_{m=0}^M\hat{a}_n|n\rangle\ee
followed by a rescaling (a filter) of the Fourier coefficients
\be |\hat{\psi}_M^R\rangle\equiv
R|\hat{\psi}_M\rangle=\sum_{m=0}^M\frac{\hat{\lambda}_n}{\lambda_n}\hat{a}_n|n\rangle\ee
renders the reconstruction formula for $\hat{\Psi}_M^R(z)=\langle
z|RP_M|\hat{\psi}\rangle$ to the expression
(\ref{reconstruccionover}). For arbitrary functions $\psi\in{\cal H}$, the new
bound for the squared normalized error turns out to be
\be \frac{\|\psi-\hat\psi^R_M\|^2}{\|\psi\|^2}\leq
\epsilon_N^2(\psi)+ \frac{\langle\psi_{M}^\perp|P_{\cal
S}P_MR^2P_MP_{\cal S}|\psi_{M}^\perp\rangle}{\|\psi\|^2}\leq
\epsilon_N^2(\psi)+\epsilon_N^2(\psi)(1+\nu_0(p,N))^2,\label{errorpsi3}\ee
where we have followed the same steps as in the proof of Theorem
\ref{maintheoremunder} in Appendix \ref{AppendixB}, used that the spectral radius
$\rho(R)=\|R\|=1+\nu_0(p,N)$ and that $RP_MP_{\cal S}P_M=P_M$.
Contrary to (\ref{errorpsi2}), the new bound (\ref{errorpsi3}) is
proportional to $\epsilon_N^2(\psi)$. If we, moreover, assume a
behavior for $a_n$ as in (\ref{asympepsilonN}), then we have that
the error (\ref{errorpsi3}) is of order $O(1/N^{2\alpha-1})$.

Let us comment on an alternative approach to the reconstruction of
functions $\psi$ for small $\epsilon_{M+1}(\psi)$,
which will turn out to be more convenient in a certain limit.
Actually, for $\epsilon_{M+1}(\psi)<<1$ we have
\be
\|\psi-P_M\psi\|^2=\epsilon_{M+1}^2(\psi)\|\psi\|^2<<\|\psi\|^2,\ee
so that, the reconstruction formula (\ref{reconstruccionover}) for
$\psi_M=P_M\psi$ would give a good approximation of $\psi$,
similarly to the approach followed in \cite{Pesenson2-1}, Section
4. The problem is that, in general, the original data
$\Psi_k=\langle z_k|\psi\rangle$ for $\psi$ and the (unknown)
``truncated'' data $\Psi_{M,k}=\langle z_k|P_M|\psi\rangle$ for
$\psi_M$ are different unless $\langle z_k|P_M=\langle z_k|,
\forall k=0,\dots,N-1$, which is equivalent to $\langle
z_k|P_M|z_k\rangle=1, \forall k=0,\dots,N-1$. The following
proposition studies the conditions under which such requirement is
approximately satisfied.

{\prop For large
$M$, the diagonal matrix elements of $P_M$ in
 ${\cal H}^{\cal S}$ have the following behavior:
\be P_M(p)\equiv \langle
z_k|P_M|z_k\rangle\simeq \left\{\ba{lcl} 1 &{\rm if}& p < p_c-{\sigma_c},\\ 0 &{\rm if}& p> p_c+{\sigma_c},\ea\right.
\label{Heaviside} \ee
(i.e., an approximation to the Heaviside (unit step)
function), where
\be p_c=M+1,\; \sigma_c=\sqrt{M+1}\label{radiocritico}\ee
denote a ``critical'' average number of particles and the standard deviation, respectively.
}

 \ni \textbf{Proof.} Using the expression (\ref{calT}) we have
\be P_M(p)\equiv \langle
z_k|P_M|z_k\rangle=\sum_{m=0}^M\ct_{km}\ct^*_{mk}=e^{-p}
\sum_{m=0}^M\frac{1}{m!}p^{m}.\ee
We can compute
\be  \frac{d P_M(p)}{d
p}=-e^{-p}\frac{p^M}{M!}=-Q_M(p).\ee
$Q_M(p)$ is an Erlang-$(M+1)$ distribution (gamma distribution for integer $M$) with average value of $p_c\equiv\int_0^\infty p Q_M(p)dp=M+1$ and variance $\sigma^2_c=\int_0^\infty (p-p_c)^2 Q_M(p)dp=M+1$. The
Erlang-$(M+1)$ distribution converges to a Normal distribution ${\cal N}(p_c,\sigma^2_c)$
for large values of $M$. Furthermore, we know that the Dirac delta distribution $\delta$ is the limit
of gaussians $\delta(x-\mu)=\lim_{\sigma\to 0}{\cal N}(\mu,\sigma^2)$. This implies that
${\cal N}(p_c,\sigma^2_c)\simeq \frac{1}{p_c}\delta(\frac{p-p_c}{p_c})$ for large values of $M$, which means
that $P_M(p)$ approaches a Heaviside function for large $M$. This is the so-called \textit{droplet} in
quantum Hall effect language (see e.g. \cite{Sakita} for an alternative proof in this context).
$\blacksquare$

\begin{figure}[h]
\begin{center}
\includegraphics[width=4.5cm,keepaspectratio]{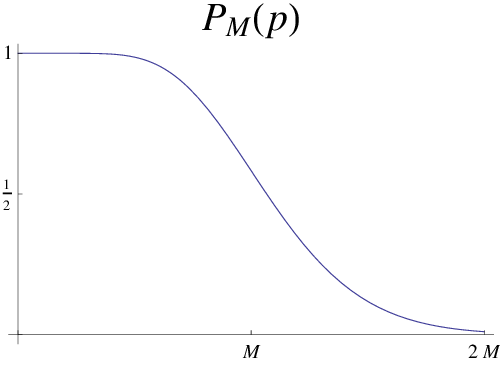}
\hspace{0.5cm}\includegraphics[width=4.5cm,keepaspectratio]{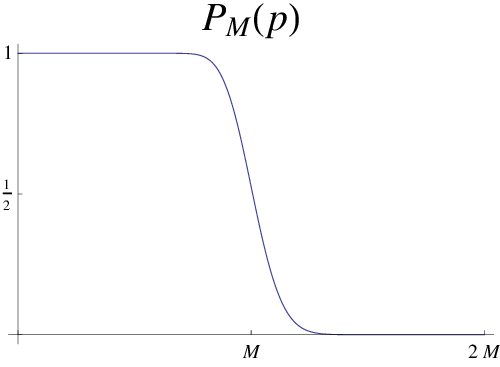}\hspace{0.5cm}\includegraphics[width=4.5cm,keepaspectratio]{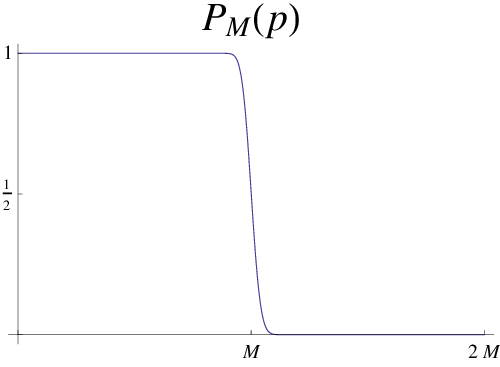}
 \end{center}
 \caption{\label{Droplet} $P_M(p)$ as a function of $p$ for three
different values of $M$: 10, 100 and 1000.}
\end{figure}

Figure \ref{Droplet} shows a plot of $P_M(p)$ as a function of
$p$ for different values of  $M$. It is clear how $P_M(p)$ approaches the step
function as $M$  grows.


{\rem The matrix elements of $P_M$ in ${\cal H}_s^{\cal S}$ have a
circulant matrix structure. In fact, they can be seen as a Fourier
transform of the coefficients $\lambda_n$. They have the
expression $\langle z_k|P_M|z_l\rangle ={\cal C}_{k-l}(p)$, where
\be {\cal C}_l(p) = \frac{1}{N}\sum_{m=0}^M \lambda_m e^{-2\pi
i\frac{m l}{N}}\,. \ee
Note that ${\cal C}_{N-l}={\cal C}_l^*$, therefore the only
independents elements are ${\cal C}_l\,,l=0,\ldots,\frac{N}{2}$.

%

\section{Conclusions and Outlook\label{comments}}

Restricting canonical CS states to the (discrete) circle provides an alternative way of analyzing functions in the Fock-Bargmann representation.
For truncated Hilbert spaces, this finite CS subsystems  provide finite tight frames with interesting properties. The lack of completeness in the general case
still allows for partial reconstructions up to a certain error which goes to zero when the number of phase samples $N\to\infty$. We also prove that an almost-exact
treatment is also possible when the number of phase samples $N$ is greater that the mean number of particles $p$ for physical states.
 An alternative exact approach would be to consider
samples of $p$ too, thus leading to ``circular-lattice'' structures.

Quantum state reconstruction is important in Quantum Optics and Quantum Computation and Information. Quantum states are reconstructed by using measurements
on an ensemble of identical quantum states. In Quantum Optics, phase measurements heavily rely on beam splitters and phase shifters to make simultaneous homodyne
measurements of the light field using photon counting statistics \cite{Photoncounting}. We feel that  our mathematical results might be of some physical usage here.

Our formalism deals with pure states, but it could be easily extended to mixed states, allowing to reconstruct accurately quasiprobability
distribution functions from discrete phase samples.

\section*{Acknowledgements}
Work partially supported by the Fundaci\'onn S\'eneca, Spanish MICINN
and Junta de Andaluc\'\i a under projects [08816/PI/08,
08814/PI/08], [FIS2008-06078-C03-01] and
FQM219, respectively. M. Calixto thanks the Fundación Séneca for a grant in the framework of ``PCTRM
2007-2010'' with funding of INFO and FEDER.

\appendix
\section{Phase operator\label{phaseopsec}}

A naive way of defining a phase operator $\Theta$ was given by
Dirac \cite{Dirac} through the polar decomposition of the
annihilation operator
\be a=e^{i\Theta}\sqrt{\cn}.\ee
The inherent difficulties with this approach (later realized by
Dirac himself) where clearly pointed out by \cite{Susskind} (see
also \cite{Nieto}). In fact, one can prove that:
\be e^{i\Theta}e^{-i\Theta}=1,\;\;
e^{-i\Theta}e^{i\Theta}=1-|0\ld\li 0|,\ee
which means that $e^{i\Theta}$ is not unitary, i.e., $\Theta$ is
nor hermitian, though it is practically so in a macroscopic system
($p>>1$). Actually, one can see that
\be \langle z|e^{i\Theta}|z\rangle=\langle
z|a\cn^{-1/2}|z\rangle=e^{i\theta}\left(1+\frac{1}{2p}\right)\simeq
e^{i\theta}\ee
in a macroscopic system ($p>>1$).

The difficulty of defining a proper (hermitian) phase operator was
circumvented in \cite{Pegg1,Pegg2}. This involves describing a
bosonic field mode in a finite (truncated) but arbitrarily large
state space ${\cal H}_M\subset{\cal H}$ of $M+1$ dimensions. The
phase states are defined as:
\be |\theta\ld=\lim_{M\to\infty}(M+1)^{-1/2}\sum_{n=0}^M
e^{in\theta}|n\ld,\ee
where the limiting procedure is necessary in order to normalize
the states and it must be taken with care (after expectation
values are calculated). Working with finite $M$, one realizes that
the phase states $|\theta_k\ld$ with
\be \theta_k=\theta_0+\frac{2\pi k}{M+1} \ee
(for arbitrary $\theta_0$) constitutes an orthonormal basis of
${\cal H}_M$. The hermitian phase operator is simply defined by
\be \Theta\equiv\sum_{k=0}^{M}\theta_k|\theta_k\ld\li\theta_k|.\ee
%

These states are, however, unphysical, in the sense that in the limit when $M\rightarrow \infty$ their mean number of particles
goes to infinity \cite{Pegg2}.

\section{Proof of Theorem \ref{maintheoremunder}}\label{AppendixB}

In this appendix we prove the main theorem \ref{maintheoremunder} of our work.
Before tackling the proof of this theorem, we shall
introduce some notation and prove some auxiliary results.

{\begin{lem} \label{lemmaunder1}  The pseudo-frame operator $\ct:{\cal H}\to \mathbb C^N$ given by
$\ct(\psi)=\{\langle z_k|\psi\rangle, z_k\in {\cal Q}\}$ [remember the construction after Eq.
(\ref{pbiop2})] is such that the overlapping kernel operator ${\cal B}=\ct\ct^*$ is an $N\times N$
Hermitian positive definite invertible matrix, admitting the eigen-decomposition ${\cal B} ={\cal
F}\hat{D}{\cal F}^*$, where $\hat{D} ={\rm diag}(\hat\lambda_0,\ldots,\hat\lambda_{N-1})$ is a diagonal
matrix with $\hat\lambda_j$ given by (\ref{lambdagorro}) and ${\cal F}$ is the standard Fourier matrix.\end{lem}}

\ni \textbf{Proof.} Let us see that ${\cal B}$ is diagonalizable
and its eigenvalues $\hat\lambda_k$ are given by the expression
(\ref{lambdagorro}), which actually shows that all are strictly
positive and hence ${\cal B}$ is invertible. This can be done by
taking advantage of the circulant structure of ${\cal B}$ (see
e.g. Appendix B in \cite{samplingsphere}). Actually, using the
expression of the CS overlap (\ref{CSoverlap}), we have:
\be {\cal B}_{kl}=\langle z_k|z_l\rangle=e^{-p}\exp\left(pe^{2\pi i(l-k)/N}\right)\equiv{\cal
C}_{l-k},\label{Bkl}\ee
where the circulant structure becomes apparent. The eigenvalues of ${\cal B}$ are easily computed by the
formula:
\be \hat\lambda_k=\hat{D}_{kk}=({\cal F}^*{\cal B}{\cal
F})_{kk}=\frac{1}{N}\sum_{n,m=0}^{N-1}e^{i2\pi kn/N}{\cal
C}_{m-n}e^{-i2\pi mk/N}.\label{lambdak} \ee
If we expand  (\ref{Bkl}) as
\be {\cal C}_l=e^{-p}\sum_{q=0}^\infty \frac{p^{q}}{q!}e^{2\pi i lq/N}\,,\ee
insert this in (\ref{lambdak}) and  use the general orthogonality
relation for Rectangular Fourier Matrices (\ref{exponencial}), we
arrive at (\ref{lambdagorro}). It is evident that
$\hat{\lambda}_k>0, \forall k=0,1,\ldots,N-1$, so that ${\cal B}$
is invertible. $\blacksquare$

Following the general guidelines of Subsec. \ref{CSandFrames}, we now introduce the
projector $P_{\cal S}$:

{\lem \label{lemmaunder2} Under the conditions of the previous Lemma, the set $\{|\tilde{z}_k\rangle
=\sum_{l=0}^{N-1} (\cb^{-1})_{lk}|z_l\rangle\,,k=0,\ldots,N-1\}$ constitutes a dual pseudo-frame for
${\cal S}$, the operator $P_{\cal S}=\ct_r^+\ct$ is an orthogonal projector onto the subspace  ${\cal
H}^{\cal S}$, where $\ct_r^+=\ct^* \cb^{-1}$ is a (right) pseudoinverse for $\ct$, and
\be
 \sum_{k=0}^{N-1} |\tilde{z}_k\rangle\langle z_k| =
\sum_{k=0}^{N-1} |z_k\rangle\langle \tilde{z}_k| =P_{\cal
S}\label{resolproy} \ee
provides a resolution of the projector $P_{\cal S}$, whose matrix elements in the orthonormal base of Fock states
(\ref{Fockbasis}) of ${\cal H}$ exhibit a structure of diagonal $N\times N$ blocks:
\be P_{mn}({p},N)\equiv\langle m|P_{\cal S}|n\rangle =(\lambda_m\lambda_n)^{1/2} \hat{\lambda}_{n
\,{\rm mod} N}^{-1}\delta_{(n-m) \,{\rm mod} N,0}, \;m,n=0,1,2,\dots,\label{projemn}\ee
with $\hat{\lambda}_n$ given by (\ref{lambdagorro}). }

\ni \textbf{Proof.} If we define $\ct_r^+=\ct^* \cb^{-1}$ it is
easy to check that $\ct \ct_r^{+}=I_N$ is the identity in
$\mathbb{C}^N$. In the same way, $P_{\cal S}=\ct_r^+\ct$ is a
projector since $P_{\cal
S}^2=\ct_r^+\ct\ct_r^+\ct=\ct_r^+\ct=P_{\cal S}$ and it is
orthogonal $P_{\cal S}^*=(\ct^* \cb^{-1}\ct)^* = \ct^* \cb^{-1}\ct
=P_{\cal S}$ since $\cb$ is self-adjoint. The resolution of the
projector is provided by Eq. (\ref{resolucionproyector}). Its
matrix elements can be calculated through:

\be P_{mn}({p},N)=\sum_{k,l=0}^{N-1}{\cal T}^*_{ml}({\cal B}^{-1})_{lk}{\cal T}_{kn}.\label{Pmn}\ee
The inverse of ${\cal B}$ can be obtained through the
eigen-decomposition:
\be ({\cal B}^{-1})_{lk}=({\cal F}\hat{D}^{-1}{\cal
F}^*)_{lk}=\frac{1}{N}\sum_{j=0}^{N-1}\hat{\lambda}_j^{-1} e^{2\pi
ij(k-l)/N}.\label{inverseB}\ee
Inserting this last expression and ${\cal
T}_{kn}=\lambda_n^{1/2}{\cal F}_{kn}$ in (\ref{Pmn}) and using the
general orthogonality relation (\ref{exponencial}) for RFM, we
finally arrive at (\ref{projemn}). $\blacksquare$

The matrix elements (\ref{projemn}) will be useful when computing the error function (\ref{errorpsi2})
for a general  function (\ref{FourExpan2}). At some point, we shall be interested in their
asymptotic behavior for large $N$ (large number of samples). In order to give an explicit expression of
this asymptotic behavior of $P_{mn}({p},N)$, it will be useful to define the following functions:
\be \nu_n({p},N)\equiv \frac{\hat{\lambda}_n-\lambda_n}{\lambda_n}=
\sum_{u=1}^{\infty}\frac{n!}{(n+uN)!}p^{uN},\; n=0,\dots,N-1.\label{varepsilon}\ee
Note that $\nu_n({p},N)=\frac{n!}{p^n}H_{N,n}(p)-1$, where $H_{N,n}(p)$ are the generalized hyperbolic functions \cite{GeneralizedHyperbolic}.
In terms of $\nu_n({p},N)$, the matrix elements (\ref{projemn}) adopt the following form:
\be P_{mn}({p},N)=\frac{(\lambda_{j+sN}\lambda_{j+tN})^{1/2}}{\sum_{u=0}^\infty\lambda_{j+uN}}
=\frac{j!}{\sqrt{(j+sN)!}\sqrt{(j+tN)!}}p^{(s+t)N/2}\frac{1}{1+\nu_j({p},N)},
\label{projemnepsilon} \ee
for $m=j+sN$ and $n=j+tN$, with $j=0,\dots,N-1$ and
$s,t=0,1,\dots$, and zero otherwise. In particular, for $n,m\leq
N-1$, the projector (\ref{projemnepsilon}) adopts the simple
diagonal form
\be P_{mn}({p},N)=\frac{1}{1+\nu_n({p},N)}\delta_{n,m},\; n,m=0,\dots,N-1.
\label{projemnepsilon1caja} \ee

Let us state and prove an interesting monotony property of $\nu_n({p},N)$.

{\lem \label{lemmaunder4} The functions (\ref{varepsilon}) are strictly decreasing sequences of $n$ for
${p}>0$, that is:
\be \nu_n({p},N)<\nu_m({p},N) \Leftrightarrow n>m,\;\; n,m=0,\dots,N-1.\label{monotonyprop}\ee
} \ni\textbf{Proof.}  The diference between two consecutive terms of the sequence is:

\[\nu_{n+1}-\nu_n=
\sum_{u=1}^{\infty}\left[\frac{(n+1)!}{(n+1+uN)!}-\frac{(n)!}{(n+uN)!}\right]p^{uN},\,n=0,\dots,N-1. \]
Moreover, it can be easily checked that
\[\frac{(n+1)!}{(n+1+uN)!}<\frac{(n)!}{(n+uN)!},\; \forall u\in \mathbb{N},
n=0,\dots,N-1\]
which implies that $\nu_{n+1}<\nu_n, n=0,\dots,N-1$ $\blacksquare$

Now we are in conditions to prove our main theorem in this section.\\

\noindent\textbf{Proof of Theorem \ref{maintheoremunder}:}
According to (\ref{discrete-repker}), the pseudo-dual frame is
defined by
\be |\tilde{z}_k\rangle = \sum_{l=0}^{N-1} ({\cal
B}^{-1})_{lk}|z_l\rangle=\sum_{l=0}^{N-1}
\frac{1}{N}\sum_{j=0}^{N-1}\hat{\lambda}_j^{-1} e^{2\pi
ij(k-l)/N}|z_l\rangle.\nn\ee
Thus, the interpolating functions (\ref{interpolating}) read:
\be L_k(z)=\langle z|\tilde{z}_k\rangle=\sum_{l=0}^{N-1}
\frac{1}{N}\sum_{j=0}^{N-1}\hat{\lambda}_j^{-1} e^{2\pi
ij(k-l)/N}\langle z|z_l\rangle.\nn\ee
Noting that
\be \langle z|{z}_l\rangle=\sum_{n=0}^{\infty}\langle z|n\rangle\langle n|{z}_l\rangle=\sum_{n=0}^{\infty}
\frac{e^{-|z|^2/2}\bar{z}^n}{\sqrt{n!}}{\cal T}^*_{ln}, \ee
and using the orthogonality relation for Fourier matrices
(\ref{exponencial}), we arrive at (\ref{hatinterpolating}).

Now it remains to prove the bound (\ref{errorpsi2}) for the error.
Decomposing $|\psi\rangle$ in terms of $|\psi_{N-1}\rangle\equiv
P_{N-1}|\psi\rangle$ and $|\psi_{N-1}^\perp\rangle\equiv
(I-P_{N-1})|\psi\rangle$, we can write
\bea {\cal E}_\psi^2({p},N)\|\psi\|^2&=&\langle \psi|\psi\rangle  -
\langle \psi_{N-1}|P_{\cal S}|\psi_{N-1}\rangle - \langle \psi_{N-1}^\perp|P_{\cal S}|\psi_{N-1}^\perp\rangle\nn  \\
& & -2 {\rm Re} \langle \psi_{N-1}|P_{\cal
S}|\psi_{N-1}^\perp\rangle. \label{errordecomp}\eea
Let us start by bounding the term
\be \langle \psi_{N-1}|P_{\cal S}|\psi_{N-1}\rangle=\sum_{n=0}^{N-1}|a_n|^2\langle n|P_{\cal
S}|n\rangle=\sum_{n=0}^{N-1}|a_n|^2\frac{1}{1+\nu_n({p},N)}\geq
\frac{1-\epsilon_{N}^2(\psi)}{1+\nu_0({p},N)}\|\psi\|^2, \ee
where we have used the expression (\ref{projemnepsilon1caja}), Lemma \ref{lemmaunder4} in bounding
$\frac{1}{1+\nu_n({p},N)}\geq \frac{1}{1+\nu_0({p},N)}, \forall n=0,\dots,N-1$ and the
definition (\ref{quasi-bandlimited}). Next we shall bound the term
\bea  -2 {\rm Re} \langle \psi_{N-1}|P_{\cal
S}|\psi_{N-1}^\perp\rangle &\leq & 2|\langle \psi_{N-1}|P_{\cal
S}|\psi_{N-1}^\perp\rangle| \leq 2\|\psi_{N-1}\| \|P_{\cal
S}\psi_{N-1}^\perp\|\nn\\ &\leq & 2\|\psi_{N-1}\| \|P_{\cal
S}\|\|\psi_{N-1}^\perp\|=
2\sqrt{1-\epsilon_N^2(\psi)}\epsilon_N(\psi)\|\psi\|^2,
 \eea
where we have used that  $P_{\cal S}$ is an orthogonal projector,
therefore its spectral radius is $\rho(P_{\cal S})=\|P_{\cal
S}\|=1$, and the Cauchy-Schwarz inequality. Using the same
arguments, we can bound the remaining term as follows:
\be -\langle \psi_{N-1}^\perp|P_{\cal
S}|\psi_{N-1}^\perp\rangle\leq \langle \psi_{N-1}^\perp|P_{\cal
S}|\psi_{N-1}^\perp\rangle\leq \|\psi_{N-1}^\perp\| \|P_{\cal
S}\|\|\psi_{N-1}^\perp\|=\epsilon_N^2(\psi)\|\psi\|^2. \ee
Putting together all this information in (\ref{errordecomp}), we
arrive at the bound (\ref{errorpsi2}) for the error.$\blacksquare$

 It can
be easily seen that $\lim_{N\to\infty} \nu_0({p},N)=0, \forall p>0$. As a consequence, the
error (\ref{errorpsi2}) goes to zero as $N\to \infty$. To obtain the order of magnitude of this error,
we shall firstly give an asymptotic behavior of $\nu_0({p},N)$ for large $N$.

{\prop \label{lemmaunder3} The quantity $\nu_0({p},N)$ has the following asymptotic behavior (as a
function of $N$):
\be
\nu_0(p,N)=\frac{1}{N!}p^{N}+O(N^{-2N}),\label{varepsilon0asymp}
\ee
as long as
\be p<
p_0(N)=\left(\frac{(2N)!}{N!}\right)^{\frac{1}{N}}=\frac{4}{e}N
\left(1+\frac{1}{2N}\ln(2)+O(\frac{1}{N^2})\right).\label{roptimo}\ee
}

\ni \textbf{Proof.} Let us see that the first addend ($u=1$) of
\be \nu_0(p,N)=
\sum_{u=1}^{\infty}\frac{1}{(uN)!}p^{uN}\label{varepsilon0}\ee
is dominant when $p<p_0(N)$. Indeed, the quotient between two
consecutive terms of the series (\ref{varepsilon0}) is
\be
p^{N}\frac{(uN)!}{((u+1)N)!}<p^{N}\frac{N!}{(2N)!},\ee
where we have used that the quotient of factorials is decreasing in
$u\in\mathbb N$. If we impose the terms of the series
(\ref{varepsilon0}) to be monotonically decreasing for any
$u\in\mathbb N$, i.e.
\be p^{N} \frac{N!}{(2N)!}<1,\ee
then we arrive at (\ref{roptimo}). Thus, the first addend, $u=1$,
of (\ref{varepsilon0}) is the leading term. Using the Stirling
formula, the asymptotic behavior of the factorial
$\frac{1}{(2N)!}$ of the second term, $u=2$, gives the
announced result (\ref{varepsilon0asymp}).  $\blacksquare$


\begin{thebibliography}{99}

\bibitem{Neumann} J. von Neumann, Mathematical Foundations of Quantum Mechanics (Princeton University Press, 1996)
\bibitem{CS} A. Perelomov, Generalized Coherent States and Their
Applications, Springer-Verlag (1986).
\bibitem{Gazeau} S.T. Ali, J-P. Antoine, J.P. Gazeau, Coherent States, Wavelets and Their
Generalizations, Springer (2000)
\bibitem{Seip} K. Seip, Density theorems for sampling and interpolation
in the Bargmann-Fock space, Bull.  Am. Math. Soc. \textbf{26} (1992) 322-328
\bibitem{Dirac} P.A.M. Dirac, The Quantum Theory of the Emission and Absorption of Radiation, Proc. Roy. Soc. (London) A114 (1927) 243-265
\bibitem{Susskind} L. Susskind and J. Glogower, Physics \textbf{1} (1964)
49
\bibitem{Nieto} P. Carruthers and M.M. Nieto, Phase and Angle Variables in Quantum Mechanics, Rev. Mod. Phys. \textbf{40}
(1968) 411-440
\bibitem{Pegg1} D.T. Pegg and S.M. Barnett, Unitary Phase Operator in Quantum Mechanics, Europhys. Lett. \textbf{6}
(1988) 483-487
\bibitem{Pegg2} D.T. Pegg and S.M. Barnett, Phase Properties of the Quantized Single-mode Electromagnetic Field,  Phys. Rev. \textbf{A39} (1989)
1665-1675
\bibitem{GazeauPLA} P.L. García de León and J.P. Gazeau, Coherent state quantization and phase operator, Physics Letters \textbf{A361} (2007) 301-304
\bibitem{KiblerJPA} M. Daoud and M. R. Kibler, Phase operators, temporally stable phase states,
mutually unbiased bases and exactly solvable quantum systems, J. Phys. A: Math. Theor. \textbf{43} (2010) 115303 (18pp)
\bibitem{GaussianStates} X.B. Wanga, T. Hiroshima, A. Tomita and M. Hayashi,
Physics Reports \textbf{448}, 1--111, (2007)
\bibitem{GazeauJPArev} N. Cotfas and J. P. Gazeau, Finite tight frames and some applications, J. Phys. A: Math. Theor. \textbf{43} (2010) 193001 (26pp)
\bibitem{pseudoinverse} A. Ben-Israel, T.N.E. Greville,
Generalized Inverses, Springer-Verlag (2003)
\bibitem{Holschneider} M. Holschneider, Wavelets: an analysis tool, Oxford University
Press (1998)
\bibitem{Fuhr} H. F\"uhr, Abstract Harmonic Analysis of Continuous Wavelet Transforms, Springer
(2005)
\bibitem{samplingsphere} M. Calixto, J. Guerrero and J.C. S\'anchez-Monreal, Sampling Theorem and Discrete Fourier Transform
on the Riemann Sphere, Journal of Fourier Analysis and Applications \textbf{14} (2008), 538-567
\bibitem{Pesenson2-1} I. Pesenson, Paley-Wiener Approximations and Multiscale
Approximations in Sobolev and Besov Spaces on Manifolds, J. Geom. Anal.
\textbf{19} (2009), 390-419
\bibitem{Sakita} B. Sakita, $W_\infty$ gauge transformations
and the electromagnetic interactions of electrons
in the lowest Landau level, Phys. Lett. \textbf{B315} (1993) 124-128
\bibitem{Photoncounting} W. P. Schleich, Quantum Optics in Phase Space, Wiley-VCH (2001).

\bibitem{GeneralizedHyperbolic}  A. Ungar, Generalized hyperbolic functions, Amer. Math. Monthly \textbf{89} (1982), 688-691
.




\end{thebibliography}
\end{document}